\documentclass[3p,times,procedia]{elsarticle}
\usepackage{amssymb}

\usepackage{amsmath}
\usepackage[utf8]{inputenc}

\usepackage[figuresright]{rotating}

\begin{document}

\begin{frontmatter}




\title{Open boundary conditions in stochastic transport processes\\
  with pair-factorized steady states}


\author[1]{Hannes Nagel}
\ead{Hannes.Nagel@itp.uni-leipzig.de}


\author[2]{Darka Labavi{\'c}}
\ead{D.Labavic@jacobs-university.de}

\author[2]{Hildegard Meyer-Ortmanns}
\ead{H.Ortmanns@jacobs-university.de}

\author[1]{Wolfhard Janke}
\ead{Wolfhard.Janke@itp.uni-leipzig.de}

\address[1]{Institut f\"ur Theoretische Physik, Universit\"at Leipzig,
  Postfach 100\,920, 04009 Leipzig, Germany}

\address[2]{School of Engineering and Science, Jacobs University
  Bremen, P.O. Box 750561, 28725 Bremen, Germany}

\begin{abstract}
  Using numerical methods
  we discuss the effects of open boundary conditions on condensation
  phenomena in the zero-range process (ZRP) and transport processes with pair-factorized
  steady states (PFSS), an extended model of the ZRP with
  nearest-neighbor interaction. For the zero-range process we compare
  to analytical results in the literature with respect to criticality
  and condensation. For the extended model we find a~similar phase
  structure, but observe supercritical phases
  with droplet formation for strong boundary drives.\\

\end{abstract}

\begin{keyword}
stochastic transport processes \sep 
pair-factorized steady states \sep 
open boundary conditions


\end{keyword}

\end{frontmatter}


\section{Introduction}
\label{sec:introduction}

Stochastic mass transport processes such as the asymmetric simple
exclusion process (ASEP) or the zero-range process
(ZRP) proposed by Spitzer (1970) 
are simple transport models
for particle hopping to improve the understanding of basic phenomena in the
dynamics of particles in driven diffusive systems. Generally these
particles are abstract and may represent objects from the microscopic
to the macroscopic scale depending on the situation and their specific
dynamics. Just as these particles and their interactions, the
underlying spatial structure is an important factor in adapting and
mapping these models to physical processes. In this work we will
consider two such processes, that feature the formation of particle
condensates in periodic systems driven far from equilibrium. We are
interested in studying these processes in a~situation, where the
system is driven by the flux of particles entering and leaving through
open boundaries. First we look into the zero-range process, where we
can compare our numerical results with analytic predictions available
from the literature. Then we discuss the effect of open boundaries on the
condensation phenomenon for an extended transport process with short-range
interactions as realized by the pair-factorized steady states (PFSS)
model introduced by Evans (2006) and Wac{\l}aw et al.~(2009a, 2009b).

\section{Zero-range process}
\label{sec:zrp}

The basic stochastic mass transport process of particle hopping consists
of a~gas of indistinguishable particles on a~one-dimensional
lattice with $L$ sites. Every site $i$ can be occupied by any number of particles
$m_{i}$. In the zero-range process, in every time step of the discrete
stochastic time evolution, a~random site $i$ is selected, where a
single particle may leave to a~neighbor with the hopping rate
$u(m_{i})$. That is, particles only interact with other particles on
the same site. The direction of the hop is determined randomly, often
with respect to rates that introduce asymmetric dynamics. For an
overview of different dynamics we refer, 
to the book by
Schadschneider et al.~(2011) 
or the review by Sch\"utz (2001).

\begin{figure}
  \centering
  (a)
  \includegraphics{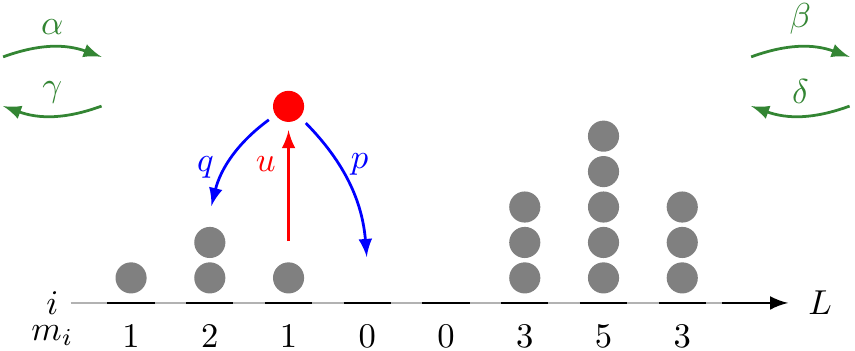}
  \hfill
  (b)
  \includegraphics{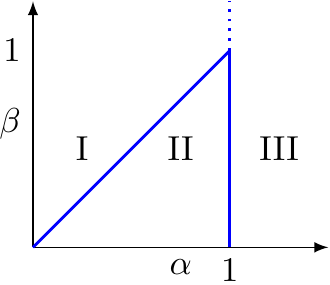}
  \caption{(a) Schematic representation of the zero-range process on a
    one-dimensional lattice with $L$ sites and particle injection
    (rates $\alpha,\delta$) and removal (rates $\gamma, \beta$)
    through open boundaries at sites $i=1,L$. (b) Phases
    induced by boundary drive in the discussed transport processes. A
    description of the phases is given in the sections of the
    respective transport process.}
  \label{fig:zrp-scheme-phases}
\end{figure}

In this work we shall consider the model
with hopping rates $u(m)=1+b/m$
on a~one-dimensional lattice with open boundary
conditions also discussed by Evans (2000) and Kafri et al.~(2002).
Under periodic boundary conditions, the main feature of this model 
is the formation of a~single-site condensate consisting of all
particles exceeding a~critical density $\rho_{\text{c}}=1/(b-2)$ for
$b>2$ such that all other sites have an average occupation equal to
$\rho_{\text{c}}$. Effects of open boundaries on this process have
been studied analytically by Levine et al.~(2005),
to which we shall first compare our results before we continue to an extended
model. Particle exchange at the boundaries at the first (last) site is
achieved by injection with rates $\alpha$ ($\delta$) and removal with
rates $\gamma$ ($\beta$), respectively, as illustrated in
Fig.~\ref{fig:zrp-scheme-phases}(a).

In the analytic study by Levine et al.~(2005), 
local fugacities are determined using a~quantum Hamiltonian
approach proposed in Sch{\"u}tz (2001) 
to find the phase
structure given in Fig.~\ref{fig:zrp-scheme-phases}(b) and respective
properties of the phases with respect to the boundary rates:
Phase~I, for $\alpha \le 1, \alpha \le \beta$, is the only phase,
where the system has a~steady state.
The total number of particles $M=\sum_{i=1}^{L} m_{i}$
remains stable, the occupation number distribution becomes
$P(m_{i}=m)=\alpha^{m}/m^{b}$ and the resulting particle density in
the bulk system is subcritical, $\rho_{\text{bulk}} <
\rho_{\text{c}}$. In phase~II, for $\alpha \le 1, \alpha > \beta$, the
rate of particle influx outweighs that of outflux and particles pile
up at the boundary site(s) before leaving, forming one or two
condensates in the totally asymmetric ($p=1, q=0$) or symmetric
($p=q=1/2$) cases, respectively. Condensation specific properties in
the bulk are not analytically determined. In phase~III, for $\alpha >
1$, the rate of particles entering at the boundary site(s) exceeds the
maximal possible hopping rate so that particles pile up on the boundary
site(s) after entering the system and the total number of particles
grows linearly in time.

We will compare only to the cases where the injection and removal
rates at each boundary are equal in the symmetric process,
$\alpha=\delta$ and $\beta=\gamma$, and $\gamma=\delta=0$ for the
totally asymmetric case. In both cases we use the hopping parameter $b=5$.

\begin{figure}
  \centering
  (a)\hspace{-4ex}
  \includegraphics{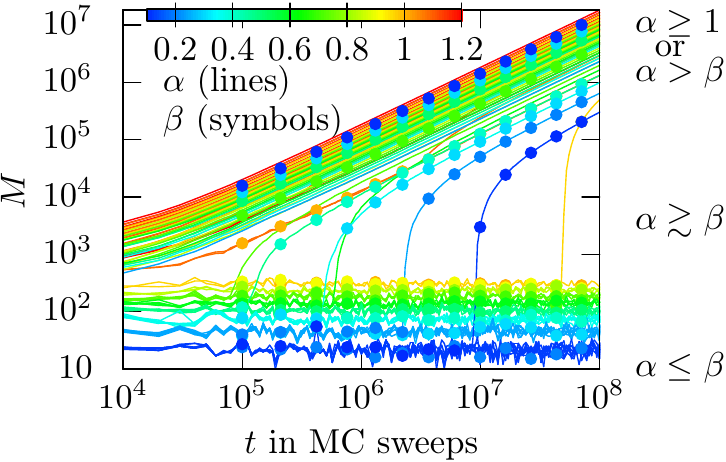}
  \hfill
  (b)\hspace{-4ex}
  \includegraphics{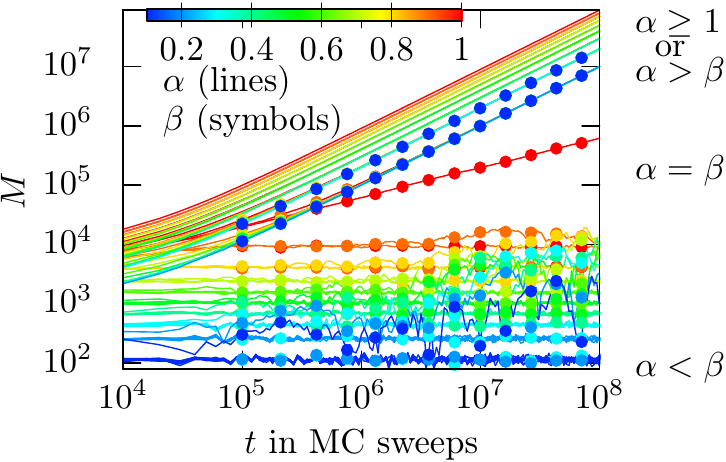}
  \caption{The time evolution of the total number of particles
    $M=\sum_{i=1}^{L} m_{i}$ for the (a) ZRP and (b) PFSS
    transport models shows the distinct differences between phases I,
    where a~steady state exists, and phases II and III. Line color
    indicates boundary rates for particle injection $\alpha$
    ($=\delta$), symbol color that of particle removal $\beta$
    ($=\gamma$). The system size is $L=1024$ sites.}
  \label{fig:zrp-totalmass}
\end{figure}

\begin{figure}[t]
  \centering (a)\hspace{-4ex}
  \includegraphics{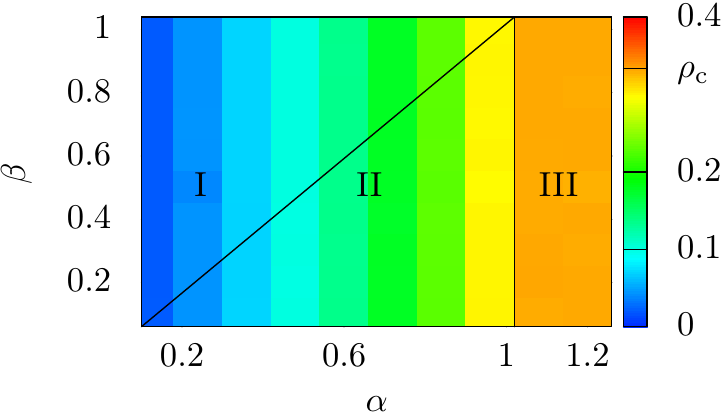}
  \hfill (b)\hspace{-4ex}
  \includegraphics{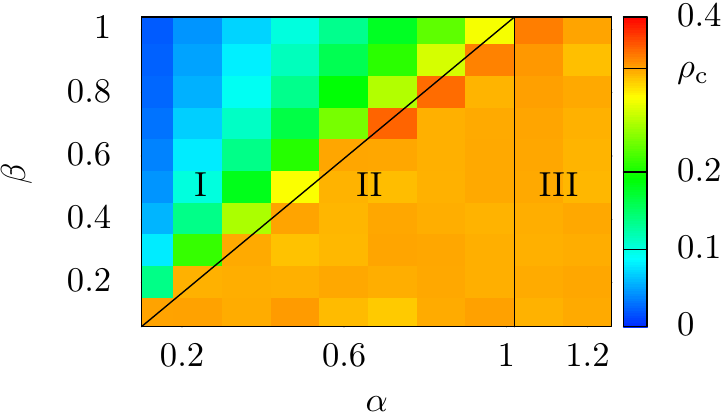}
  \caption{Bulk densities $\rho_{\text{bulk}}$ of the (a) totally
    antisymmetric and (b) symmetric ZRP with $\alpha=\delta$, 
    $\beta=\gamma$  and hopping parameter $b=5$ averaged over time for various
    values of boundary rates $\alpha$, $\beta$. The black
    lines indicate the phase transition lines. In phase III of the
    totally asymmetric case and phases II and III of the symmetric
    case, the bulk density becomes equal to the critical density
    $\rho_{\text{c}}=1/3$. The system size is $L=1024$ sites.}
  \label{fig:zrp-bulk-densities}
\end{figure}

By numerical simulation these phases and properties are easily
reproduced and visible in observables such as the total number of
particles $M$ over time as shown in
Fig.~\ref{fig:zrp-totalmass}(a) with the same phase boundaries as
predicted.
%
A more detailed look at the criticality of the system with respect to
the boundary drive is possible by computing the bulk density deep
inside the system as shown in Fig.~\ref{fig:zrp-bulk-densities} for
the (a)~totally asymmetric and (b)~symmetric processes. In the totally
asymmetric case, the condensate on the last site in phase II cannot
contribute to the bulk density which remains low and subcritical. In
phase III, a~particle condensate forms at the first site acting as a
reservoir for the bulk of the system that becomes critical
($\rho_{\text{bulk}}=\rho_{\text{c}}=1/(b-2)=1/3$) as it acts like the
bulk in a~periodic system with a~condensate. In contrast, in the
symmetric system in both phases II and III these condensates form at
the two boundary sites and particles may hop back into the system, so
that the bulk becomes critical already for $\alpha > \beta$ in
phase~II. Very small droplets of several particles become visible in
the bulk, but the monotonic falloff of the distribution of occupation
numbers in the phases~II and~III confirms the absence of condensates in the
bulk.

\section{Stochastic transport with nearest-neighbor interactions and
  pair-factorized steady state}
\label{seq:pfss}

An interesting extension of the ZRP is the introduction of
nearest-neighbor interactions: spatially extended condensates are
observed for periodic lattices in the model proposed by Evans et
al.~(2006), 
where the condensation process is qualitatively
similar to that of the zero-range process. We will consider a~slightly different 
model with hopping rates
\begin{equation}
  u(m_{i}\vert m_{i-1}, m_{i+1}) = 
  \frac{g(m_{i}-1,m_{i-1})}{g(m_{i},m_{i-1})}
  \frac{g(m_{i}-1,m_{i+1})}{g(m_{i},m_{i+i})}
  ,\; \text{where}\; g(m,n)= \sqrt{ \text{e}^{-m^{c}}
  \text{e}^{-n^{c}} } \text{e}^{-\vert m - n \vert^{b}}
  \label{eq:pfss-generic-rate}
\end{equation}
is a~two-point weight function that allows the tuning of the critical
density as well as the condensate's width and shape
between single-site, rectangular and smooth bell-like shapes as discussed by Wac{\l}aw et al.~(2009a, 2009b) and Ehrenpreis et al.~(2014). Instead
of a~fully factorized steady state as in the ZRP, this extended model
has a~pair-factorized steady state (PFSS).  For an overview of the
condensate shapes and other properties see Ehrenpreis et al.~(2014).
We choose this model to be able to
also study the effects of open boundaries in the transition region
between ZRP-like single-site condensates and extended shapes. However,
in this work we first concentrate on a~single specific
parameterization, $b=1.2, c=0.6$, where we know from previous
work by Wac{\l}aw et al.~(2009a, 2009b) and Ehrenpreis et al.~(2014)
that the periodic system exhibits smooth bell-like condensate
shapes. The specific parameter point is chosen because of the
comparably fast dynamics, low critical density and intermediate
condensate extension.

Again, we can recognize the rough phase structure by determining the
total number of particles $M=\sum_{i=1}^{L} m_{i}$ in the system versus time and
get a~similar picture as for the ZRP in
Fig.~\ref{fig:zrp-totalmass}(b). For $\alpha < 1$, $\alpha < \beta$,
$M$ remains stable and linear growth of $M$ is observed for $\alpha >
\beta$ as well as for $\alpha \ge 1$. Only directly at the transition
between the phases~I and~II, for $\alpha=\beta$, an increase in the square
root of time, $M(t)\sim \sqrt{t}$, is observed. Similar to the
behavior of the ZRP near this transition line ($\alpha \gtrsim
\beta$), for lower boundary rates $\alpha=\beta=\gamma=\delta$ the
system may remain in the steady state of phase~I for a~long time with
a stable total particle number and then suddenly jumps to the observed
behavior. 
For $\alpha > \beta$ as well as $\alpha \ge 1$ linear growth is
observed due to particles piling up on the boundary sites as for the
ZRP.

\begin{figure}[t]
  \centering
  (a)\hspace{-4ex}
  \includegraphics{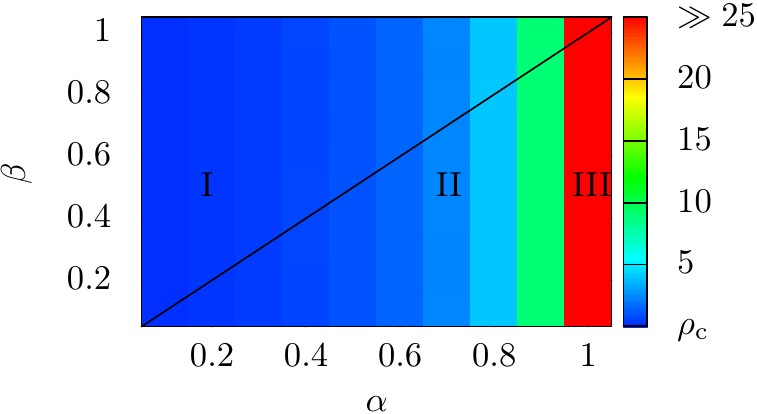}
  \hfill
  (b)\hspace{-4ex}
  \includegraphics{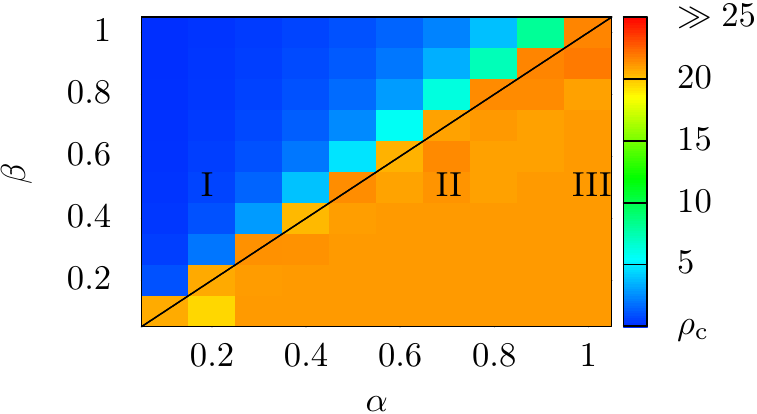}
  \caption{Bulk densities $\rho_{\text{bulk}}$ of the (a) totally
	  asymmetric and (b) symmetric PFSS (with $\alpha=\delta$, 
	  $\beta=\gamma$) transport processes averaged over time for
	  various values of boundary rates $\alpha$, $\beta$.
    Here the bulk densities become much larger than the
    critical density $\rho_{\text{c}}\approx 0.11$ in phase III
    (totally asymmetric case) and phases II and III (symmetric case),
    respectively. The system size is $L=1024$ sites.}
  \label{fig:pfss-bulk-densities}
\end{figure}

To determine the effect of these boundary condensates on the state and the
criticality
 of the bulk system, we computed the density
$\rho_{\text{bulk}}$ deep inside the system, droplet mass distributions
and average occupation number profiles. The structure of the obtained
bulk densities for various boundary rates as shown in
Fig.~\ref{fig:pfss-bulk-densities} easily compares with that of the
ZRP in Fig.~\ref{fig:zrp-bulk-densities}. The phase structure is
again visible looking at the totally asymmetric
(Fig.~\ref{fig:pfss-bulk-densities}(a)) and symmetric
(Fig.~\ref{fig:pfss-bulk-densities}(b)) models. However, the average
bulk density in phases II and III is clearly above the critical
density $\rho_{\text{c}}\approx 0.11$ as determined in Ehrenpreis et al.~(2014)
for the model
on a~periodic lattice, where rapidly emerging particle condensates
would be expected. We also observe an increase of the average bulk
densities with the square root of simulation time in phases II and
III, while it is stable in the steady-state phase I. This growth in
the bulk system can be easily understood when the system has a~long-range
correlation between sites deep inside the system and the boundary
sites. In fact, we observe such long-range correlations in the average
occupation numbers
at 
one
boundary site for the totally asymmetric process and at 
both
boundary sites for the symmetric process.




\begin{figure}[t]
  \centering (a)\hspace{-4ex}
  \includegraphics{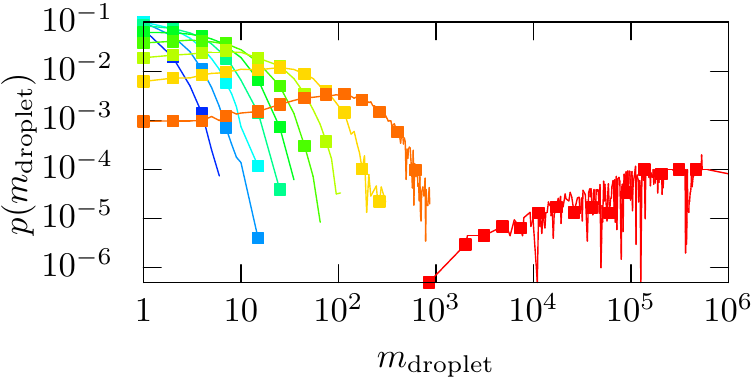}
  \hfill (b)\hspace{-4ex}
  \includegraphics{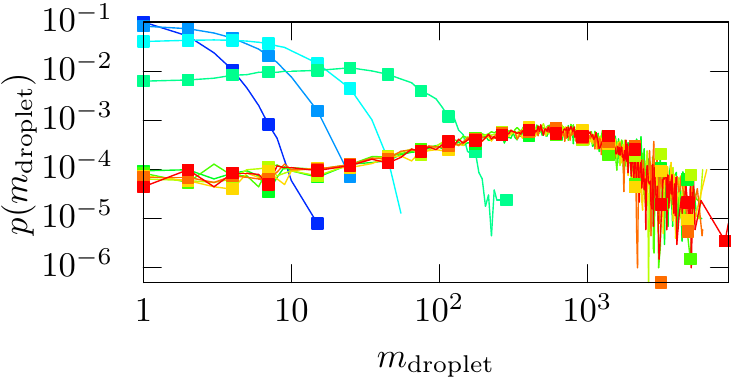}
  \\
  \includegraphics{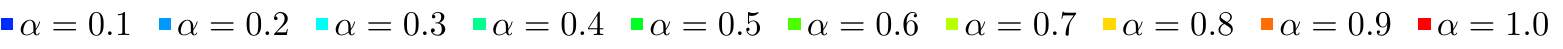}
  \caption{Droplet size distributions for the (a) totally asymmetric
    and (b) symmetric PFSS processes. Parameters range from
    $\alpha=0.1$ (blue) to $\alpha=1.0$ (red) with $\beta=0.5$
    constant%
    . For better readability the
    distributions have been binned.}
  \label{fig:sympfss-droplet-dist}
\end{figure}

The values of the bulk density above the critical density of the
periodic system raise the question whether particle condensates are
formed in the bulk system. To answer this question we first
concentrate on the distribution of droplet masses in the bulk. In a
system without condensation, this distribution is expected to fall off
monotonically, while a~bump in the tail indicates an excess of larger
droplets as a~product of the condensation process. For the totally
asymmetric process (Fig.~\ref{fig:sympfss-droplet-dist}(a)) a
shift to larger droplet masses for increasing $\alpha$ in phases I and
II due to an increasing bulk density is observed. The cause of the
distinctively different distribution for $\alpha=1$ in this figure is
that the bulk does not contain separated droplets but consists of
highly occupied sites only. In the symmetric process (Fig.~\ref{fig:sympfss-droplet-dist}(b)) a~similar shift to larger
droplet masses is observed only in phase~I. For $\alpha\ge\beta$ the
droplet mass distributions then collapse onto each other with a~bump
for large droplets. This is a~strong indicator that indeed a
condensation process takes place in the bulk of phases~II and III.
 Since, in these
phases, the number of particles in the system as well as the bulk
density grow in time, this condensation process is likely stationary.
That is, the droplets keep growing while the system evolves.
In both cases, totally asymmetric and symmetric, a~bump 
in the
droplet mass distribution develops just before the transition lines ($\alpha
\ge 1.0$ and $\alpha \ge \beta$, respectively) are crossed. We
interpret these as finite-size effects due to the limited distance of
the bulk to the boundary sites.
%

\section{Summary}
\label{sec:summary}

We studied effects of 
open boundaries on 
condensation
phenomena in the zero-range process (ZRP), reproducing the analytic
results in Levine et al.~(2005), 
and an extended
transport process~\eqref{eq:pfss-generic-rate} with nearest-neighbor
interactions. We found that boundary drive in the pair-factorized 
steady states (PFSS) model creates mostly
the same phase structure as in the ZRP, with distinct behavior
directly on the transition line between phases I and II. However,
different properties are observed in phases II and III, where no
steady state exists for the ZRP. Due to effective long-range
interactions, the boundary drive is still visible deep inside the bulk of
the system, increasing the particle density
above 
its critical value,
thus creating many large droplets but no well separated or
even single condensates as in the periodic lattice model.


As an outlook to further research into this system, we would like to
study specifically whether the tunable extended PFSS
model~\eqref{eq:pfss-generic-rate} in the single-site condensate
regime $b<c$ maps onto the ZRP with respect to boundary
drive effects and how the transition to extended rectangular and
smooth bell-like shaped condensates is reflected by the
discussed properties of the boundary drive.
More specifically, does the sensitivity to the boundary depend on the shape of the extended condensate in the case of the PFSS model?
It would also be of interest to elucidate how the lattice size influences the size of
formed droplets via the time it takes for
droplets to move to the boundary.

%
%


\section*{Acknowledgements}
We would like to thank the DFG (German Science Foundation) for
financial support under the twin Grants No.\ 
JA~483/27-1 
and
ME~1332/17-1. 
We further acknowledge support by the DFH-UFA
graduate school 
CDFA-02-07.



  


\makeatletter
\renewcommand\@biblabel[1]{}
\makeatother 


\enlargethispage{2ex}


\end{document}